\documentclass[conference]{IEEEtran}

\newtheorem{lemma}{Lemma}

\newtheorem{theorem}{Theorem}

\usepackage[dvips]{graphics}
\usepackage{epsfig}
\usepackage{amsmath}
\usepackage{amsopn}
\usepackage{amssymb}
\usepackage{color}

\begin{document}

\title{Competition and Request Routing Policies in Content Delivery Networks}

\author{\authorblockN{S. M. Nazrul Alam}
\authorblockA{Department of Computer Science\\
Cornell University\\
Ithaca, NY 14850 USA\\
Email: smna@cs.cornell.edu}
\and
\authorblockN{Peter Marbach}
\authorblockA{Department of Computer Science\\
University of Toronto\\
Toronto, ON M5S 3G4 Canada\\
Email: marbach@cs.toronto.edu}}

\maketitle

\begin{abstract}
The role of competition and monetary benefits in the design of Content Delivery Networks (CDNs) is largely an unexplored area. In this paper, we investigate the effect of competition among the competitive web based CDNs and show that little difference in their performance may cause significant financial gain/loss. It turns out that the economy of scale effect is very significant for the success of a CDN in a competitive market. So CDN peering might be a good idea. Since performance and conforming to the service level agreement (SLA) with content providers is very important, we then focus on designing CDN from this perspective. We provide an asymptotically optimal static request routing policy for a CDN under a model where a CDN company guarantees a certain level of user latency to the content providers in the SLA.
\end{abstract}

\IEEEpeerreviewmaketitle

\section{Introduction}
Content distribution (or, delivery) networks (CDNs) provide a means to improve the performance of web-based 
applications where clients access information from an origin server~\cite{Verma:2002}. A CDN consists of a set of 
surrogate servers that are placed at different ``points'' in the Internet. By directing requests to surrogate servers 
that are located close (in terms of latency) to clients, CDNs avoid congested paths (parts) of the network and thus 
are able to significantly reduce the response time of web-based applications.

In principle, each web-based content provider could setup its own CDN to better serve its clients. However, for most %%@
content providers it is financially and technically not feasible to maintain their own CDN. Therefore, in practice, %%@
content providers contract a CDN company to provide this service~\cite{akamai}~\cite{mirrorimage}. 
Typically, a CDN company and a  content provider sign a service level agreement (SLA) which determines the %%@
performance (quality of service) and price of the service provided by the CDN company to the content provider.

In this paper, we are interested in how economic aspects of commercial CDN companies influence the design of network %%@
protocols for CDNs (such as how client requests are optimally routed to surrogate servers). Like any other 
business organization, the goal of a rational commercial CDN company is  
maximizing its monetary benefit. There are two main factors which affect the monetary benefit of a CDN company: the %%@
performance and the price of the service. Naturally, when two CDN companies charge the same price then a content %%@
provider will contract the CDN which offers the better performance. On the other hand, if two CDN companies offer the %%@
same performance, content providers will choose the one with the lower price. Therefore, in order to predict the %%@
monetary benefit, or optimize the operation of a CDN company, one needs to develop an economic model that relates the %%@
performance and monetary benefit in the situation where several CDN companies compete for market share. To the best %%@
of our knowledge, there is no previous work on protocol design for CDN to maximize the monetary benefit in a %%@
competitive market.

For our analysis, we consider the situation where the service level 
agreement (SLA) guarantees a certain level of performance improvement $\beta$ (given by the ratio of new and old %%@
request response time for clients) for the content provider. For the situation where a few CDN companies compete in a %%@
large market (in terms of content providers), we show that the CDN company which offers the best performance (in %%@
terms of $\beta$) will dominate the market. For example, in the case of two competitive CDNs, our model shows that 
if one CDN has a very slight performance
 advantage over its rival CDN (i.e., a better value of $\beta$) 
then in equilibrium its revenue is at least four times higher than its rival. It is worth noting that this result is %%@
in line with what is happening in practice. Note that a CDN company can improve its performance by (a) deploying a %%@
large number of surrogates server and (b) deploying sophisticated algorithms for optimally routing client requests to %%@
surrogate servers. The above result then implies that large CDN companies (in terms of surrogate server) with %%@
sophisticated algorithms have a huge competitive advantage and will dominate the market. And indeed, currently the %%@
CDN market is dominated by a (very) small number of large CDN companies that closely guard their algorithms for %%@
client requests routing in order to keep their competitive advantage.

Having established that performance (with respect to the SLA) is of utmost importance, we next develop a request %%@
routing (or, redirection) policy to maximize the number of requests that are served within the agreed value of %%@
$\beta$ in the SLA. In particular, we propose a simple  static  request routing policy and show that it is %%@
asymptotically optimal. 

The rest of the paper is organized as follows. Section~\ref{sec:comp} describes the relationship between the %%@
performance and the revenue of web-based CDNs in 
a competitive market.
Section~\ref{sec:routingpolicy} describes the existing request routing policies
and presents a model for routing policies that is based on service level agreement. A static request routing policy %%@
for this model is proposed and it has been proven that this policy is asymptotically optimal. 
Numerical results are given in the corresponding sections.
Section~\ref{sec:conclusion} concludes 
the paper and suggests future research directions.

\section{Relationship between Performance and Revenue of a CDN in a Competitive Market}
\label{sec:comp}
In the case of web-based competitive content delivery networks, 
our goal is to understand how the
performance of a CDN affects its competitiveness. 
Doing an exact analysis to get the equilibrium point requires solving a system of 
linear equations where the number of equations 
increases with the number of CDNs involved. As a result, although numerical calculation of equilibrium revenues 
for particular values of various parameters are relatively easy for arbitrary number of competitive CDNs, %%@
theoretically
proving the general relationship among the equilibrium revenues is 
 difficult for the general case.
So instead of modeling a general case involving arbitrary number of CDNs,
we use a simple model of two CDNs  
to understand the relationship between the performance 
and the revenue of a CDN in the competitive
environment. Later we extend the model and analyze the competition among 
three CDNs. In both cases, it is found that in equilibrium 
a CDN having better performance attracts 
substantially more users at a higher price 
and thus generates significantly more revenue than
its competitors.   

Most of the time, the financial fate of
a CDN is determined by the service level agreement (SLA) signed between
CDN and the content provider. Service Level Agreements are documents that specify exactly what services the service %%@
provider will provide,
how much the customer will pay, and what will happen when things go wrong~\cite{Kaye:2002}. 
Since a typical content provider is a small entity and lacks 
necessary expertise and resources to  
negotiate the terms and conditions of a SLA with a CDN, usually
a CDN determines the terms and conditions of its own SLA for all
of its potential customers and then publishes it. A content provider
looks at the SLAs of different CDNs and then chooses a CDN based
on the services, prices and penalties written on the SLA
so that its utility is maximized. The SLAs of different CDNs may vary but
all of them should have something in common, e.g., what would be the guaranteed 
minimum improvement of user latency, what would be the level of price charged by the
CDN for the service and how much penalty the CDN would have to pay to the 
content provider when it is unable to achieve the guaranteed performance.
Usually, a CDN does not get extra money if its level of performance is higher than the
level guaranteed in the SLA. Since most content providers themselves are unable to 
monitor the performance, sometime independent 
third party companies monitor the performance of the CDNs.

\subsection{The Model}
Our model has some similarity with the model of~\cite{Gibbens:2000}, which they use to understand the competition for %%@
Internet service classes. 
In our model, we use a performance parameter $\beta$ to quantify the performance of a CDN and 
a sensitivity parameter $\theta$ to quantify the demand of a content provider. 

\subsubsection{Model of the CDNs}
Quantifying the performance of a CDN is very difficult. Traditionally, 
lower user latency and higher availability have been the two most important services provided
by a CDN. Since most CDN usually guarantees 100\% availability, availability should not be 
a big issue while analyzing the competition. Recently, content delivery networks
are offering various other services such as web hosting. However, for simplicity, in our model
we use user latency as the sole indicator of the performance of a CDN. 

From the 
point of view of user latency, the performance of a CDN can be measured by the improvement
in user latency. Let, $\psi'$ be the time required to serve a typical request from an origin server and 
$\psi_k$ be the time needed to serve the same request 
when a CDN say, $CDN_k$, serves it. In other word, $\psi'$ is the old user
latency and $\psi_k$ is the new user latency when $CDN_k$ serves the request.  
Then the performance of $CDN_k$ for that request can be defined as 
$$\beta_k = \frac{\psi_k}{\psi'}.$$
The minimum (best) possible value of the $\psi_k$ is zero and the maximum (worst)
value of $\psi_k$ should be equal to the value of $\psi'$. The worst case scenario
occurs when a request is served by the origin server and then the value of $\beta_k$ is 1.
 No rational 
content provider would hire any $CDN_k$ having $\beta_k\geq 1$.  
So we assume the value of $\beta$ lies between 
0 and 1 for all CDN. The smaller the value of $\beta$, 
the better the performance of the corresponding CDN.

Each $CDN_k$ announces its performance parameter $\beta_k$ and a price $w_k$ to the content providers. A 
content provider chooses a CDN based on the values of $\beta_k$ and $w_k$ with an objective to maximize its own %%@
utility.
In this model, it is assumed that $\beta_k$ of each $CDN_k$ is fixed but the CDN can change  its price $w_k$ to %%@
increase its competitiveness. The ultimate goal of a CDN is assumed to maximize its revenue. 

The revenue function of a CDN is defined as follows. Let, there are $K$ CDNs, say $CDN_1, CDN_2,...,CDN_K$, having
performance parameters $\beta_1, \beta_2,...,\beta_K$, respectively. They compete among them by changing their
respective prices $w_1, w_2,...,w_K$ to maximize their revenues.
Let, $\Lambda_1(w_1, w_2,...,w_K)$, $\Lambda_2(w_1,w_2,...,w_K)$,$...$,$\Lambda_K(w_1,w_2,...,w_K)$ be the number of %%@
customers (content providers) of $CDN_1, CDN_2,...,CDN_K$, respectively. The number of content providers of a CDN %%@
depends on its price as well as the price of its competitors. Definitely, $\Lambda$s  depend on the values of %%@
$\beta$s as well, but  for simplicity we keep that implicit, and note that the values of $\beta$ of each CDN is fixed %%@
during the whole competition process. 
Since the revenue of a CDN can be obtained by multiplying the  
number of customers with price, the revenue of $CDN_k$ can be defined as $$J_k(w_1, w_2,...,w_K)= %%@
\Lambda_1(w_1,w_2,...,w_K) \times w_k .$$ 

Note that all CDNs would change their prices until they reach a point where a unilateral change of price
does not increase the revenue of that CDN. Such a  point is usually refer to as a Nash Equilibrium. We actually want %%@
to find the
revenues at Nash Equilibrium. A Nash Equilibrium can be formally defined as follows.

A Nash Equilibrium is a set of prices $(w_1^*, w_2^*,...,w_k^*,...,w_K^*)$, such that for any CDN, say $CDN_k$,
for all values of $w_k$, 
$$ J_k(w_1^*, w_2^*,...,w_k^*,..,w_K^*) \geq J_1(w_1^*, w_2^*,...,w_k,..,w_K^*).$$

In words: in a Nash Equilibrium, no CDN has a unilateral incentive to change its
strategy (i.e., price). 

From now on, we would write simply  $\Lambda_k$ and $J_k$ instead of $\Lambda_k(w_1, w_2,...,w_K)$ and  %%@
$J_k(w_1,w_2,...,w_K).$

\subsubsection{Model of the Content Providers}
All content providers are not equally sensitive to the service provided by a CDN. The benefit due to lower user %%@
latency
largely depends on the type of the content and the end users of the content provider. We use a sensitivity parameter %%@
$\theta$ to represent this
sensitivity of the content providers. 

We can define $\theta$ as the maximum utility that can be increased by any CDN for that particular content provider.
From the property of $\beta$, we know that the best possible value of $\beta$ is zero and the worst possible value is %%@
1. So when $\beta_k=1$, 
there is no gain in utility; and when $\beta_k= 0$ the gain in utility is maximum and it is $\theta-w_k$, where $w_k$ %%@
is the price charged by $CDN_k$.
So, when a content provider with sensitivity parameter $\theta$ hires $CDN_k$, the pay off function can be 
defined as 
$$U(\theta, k)= \theta \times (1-\beta_k)-w_k.$$

In our model, all content providers are rational and they select the CDN which has the maximum value for the pay off %%@
function, i.e.,
a content provider with sensitivity $\theta$ selects $CDN_j$ if and only if
$$U(\theta, j) = \max_{\forall k} U(\theta, k).$$

\subsubsection{Assumptions}

Other important assumptions of our model are stated below:
\begin{itemize}

\item The utility of a content provider increases linearly with the decrease of
$\beta$ in the interval [0,1]. Although this assumption might not be the case
for some content providers, it captures the relevant feature that utility increases
with the decrease of $\beta$ while maintaining tractability.

\item No content provider hire more than one CDN. It is a reasonable and practical 
assumption.

\item Content providers differ in their preferences for lower user latency. So different
content providers have different values for the sensitivity parameter $\theta$. To reflect
the range of preferences in the population of content providers in the simplest
manner, it is assumed that there is a continuum of content providers whose $\theta$ 
parameters form a population distribution which is uniformly distributed on the interval [0,1],
and the number of content providers $\Lambda$ is a very big number. 
The uniform distribution is commonly used in 
Economics for modeling competition among firms whose
products are of different qualities; see \cite{Shaked:1982, Gabszewicz:1986,Tirole:1988}
among many others.
This assumption tells that the total number of content providers having sensitivity between $\theta_1$
and $\theta_2$ (where $0 \leq \theta_1 \leq \theta_2 \leq 1$) is $\Lambda (\theta_2-\theta_1)$. We make an additional 
assumption that the number of requests served by each CDN is proportional to the number of content providers it has. %%@
That
means that in order to compare the revenues of competitive CDNs, 
we need to take into account only their price and the number of content providers they have. Real life scenario where %%@
content providers vary in their amount of content can be simulated in this model by representing a content provider %%@
with a number of content providers which is proportional to its content size.  

\item CDNs have no capacity constraints.  That is each CDN can serve as many content providers as it can get. This is %%@
actually not very unrealistic assumption. If a CDN can not serve a request, it can always redirect it to the origin %%@
servers. In that case, the performance parameter $\beta$ of that CDN has a worse value. 
\end{itemize}
A strategy for a content provider is a choice of CDN to join, given the prices quoted by the
CDNs. Throughout this paper, we would use a superscript $*$ to denote various terms at equilibrium (e.g., at %%@
equilibrium
the price, the number of content providers
and the revenue of $CDN_1$ are $w_1^*$, $\Lambda_1^*$, $J_1^*$, respectively).

\subsection{Competition between two CDNs}
In this scenario, two CDNs, say $CDN_1$ and $CDN_2$, having performance parameter 
$\beta_1$ and $\beta_2$, charges price $w_1$ and $w_2$, respectively.
Let $J_1(w_1, w_2)$ and $J_2(w_1, w_2)$ be the revenue of $CDN_1$ and $CDN_2$, respectively.
 
Now the question we want to answer is:
\begin{quote}
Given the performance parameters $\beta_1$ and $\beta_2$, what is the relation between the equilibrium revenues of %%@
$CDN_1$ and $CDN_2$ in a duopoly?
\end{quote}

For the sake of analysis, without loss of generality we assume $\beta_2\geq \beta_1$. In order to find the revenues %%@
of the two CDNs
at equilibrium at first we need to prove a few lemmas.

\begin{lemma}
\label{lem:property1}
If $CDN_2$ is rational and $\beta_2 \geq \beta_1$, then $CDN_2$ would always set price such that $w_2 \leq w_1.$
\end{lemma}
\begin{proof}
The proof is straightforward. If $w_2 > w_1$, 
then for any content provider, pay off function for $CDN_1$ has higher value than that of the payoff function for %%@
$CDN_2$. 
So no rational content provider would choose $CDN_2$ 
and its revenue would be
zero. Since the objective of $CDN_2$ is to maximize its revenue, it will never charge price
higher than that of $CDN_1$.  
\end{proof}

When the value $\beta_1$ is equal to that of $\beta_2$, analysis is much easier;
both CDNs essentially end up with charging
same price and each gets the same number of content providers. So we focus on the case where 
$\beta_2 > \beta_1$. In this case, 
$w_2$ must be less than $w_1$, otherwise no rational content provider would hire $CDN_2$.

A content provider's selection of a particular CDN would depend on its sensitivity parameter
$\theta$. Following lemma shows best choices for different values of $\theta$. 

\begin{lemma} 
\label{lem:property2}
There exist $\theta_1$ and $\theta_2$ such that for a content provider with sensitivity parameter $\theta$, if  
$0 \leq \theta \leq \theta_2$ then $0 \geq U(\theta, 2)$ and $0 \geq U(\theta,1),$  i.e., choosing no CDN is the best %%@
option;
if  $\theta_2 \leq \theta \leq \theta_1$ then $U(\theta, 2) \geq U(\theta,1)$ and  $U(\theta,2) \geq 0,$ i.e., %%@
choosing $CDN_2$ is the best option;
if  $\theta_1 \leq \theta \leq 1$ then $U(\theta, 1) \geq U(\theta,2)$ and $U(\theta,1) \geq 0,$ i.e., choosing %%@
$CDN_1$ is the best option.
\end{lemma}

\begin{proof}
Fig.~\ref{fig:infprice} describes the constraints of the CDNs in choosing their prices. Since $\beta_1$ and $\beta_2$ %%@
are fixed, the slope of the utility
curve for each CDN is fixed. The CDNs can only change the prices to change the y-intercept. A content provider %%@
chooses the CDN which has the higher
utility at its sensitivity level. If the utilities of both CDNs are negative, then the content provider does not %%@
choose any CDN.  Fig.~\ref{fig:infprice}
shows two scenarios where at least one CDN has no revenue, so the price charged by that CDN is not a feasible price %%@
at all. Fig.~\ref{fig:feaprice} shows 
the feasible scenario where both CDNs have nonzero revenues and this lemma follows from this figure.
\end{proof}

The two CDNs would play a non-cooperative game to maximize their own revenue. 
In Nash-Equilibrium, both
CDNs would settle in prices $w_1=w_1^*$ and $w_2=w_2^*$ such that any unilateral change of
price would not increase the revenue. We consider that point as the equilibrium point. 
Following theorem shows that in equilibrium, the revenue of $CDN_1$ is significantly higher
than that of $CDN_2.$

\begin{theorem}
\label{thm:tworevenue}
In equilibrium total revenue of $CDN_1$ is at least four times of that of $CDN_2$, i.e., we have,
$$J_1^* > 4 J_2^*.$$
\end{theorem}
\begin{proof}

The critical value of $\theta_1$ is the identity of marginal content provider who is indifferent between $CDN_1$ and %%@
$CDN_2$, so we have
$U(\theta_1, 1)=U (\theta_1, 2),$ which implies that 
$\theta_1\times(1-\beta_1)-w_1 = \theta_1\times(1-\beta_2)-w_2$, i.e.,
$$\theta_1 = \frac{w_1 - w_2}{\beta_2 - \beta_1}.$$
Again, the critical value of $\theta_2$ is the sensitivity of the marginal user who is 
indifferent between hiring $CDN_2$ and hiring no CDN at all, so we have
$U(\theta_2,1)=0,$ which implies that
$\theta_2\times(1-\beta_2)-w_2 = 0$, i.e.,
$$\theta_2=\frac{w_2}{1-\beta_2}.$$
The content providers having the value of their sensitivity between $\theta_1$ and 1 will choose
$CDN_1$ (see Fig.~\ref{fig:feaprice}), while the content providers having the value of their 
sensitivity between $\theta_2$ and $\theta_1$ will choose
$CDN_2$. So the masses
of content providers who join $CDN_1$ and $CDN_2$ are $(1-\theta_1)\times\Lambda$ and %%@
$(\theta_1-\theta_2)\times\Lambda,$  respectively
 (given  the assumption that the preferences are uniformly distributed on the unit interval
 and $\Lambda$ is the total number of content providers).

Now, total revenue of $CDN_1$ is
$J_1 = (1-\theta_1)\times\Lambda \times w_1$, i.e.,
$$J_1 = \left(1-\frac{w_1 - w_2}{\beta_2 -
\beta_1}\right)\times \Lambda\times w_1.$$
Similarly, total revenue of $CDN_2$ is 
$J_2 = \left(\theta_1- \theta_2\right)\times\Lambda\times w_2$, i.e.,
$$J_2 = \left(\frac{w_1 - w_2}{\beta_2 -
\beta_1}- \frac{w_2}{1-\beta_2}\right)\times\Lambda\times w_2.$$
To get the point where revenue is maximum, 
set first derivative to zero (provided the second derivative
is negative).

\begin{figure}
\begin{center}
\input{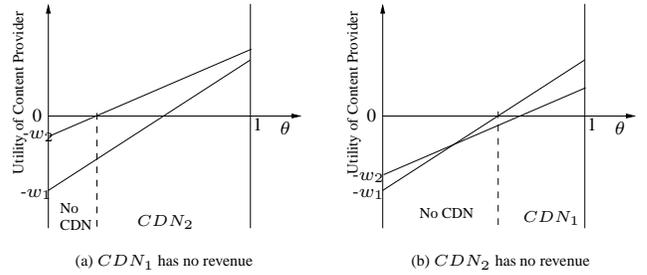}
\caption{Price settings which are infeasible.}
\label{fig:infprice}
\end{center}
\end{figure}

\begin{figure}
\begin{center}
\input{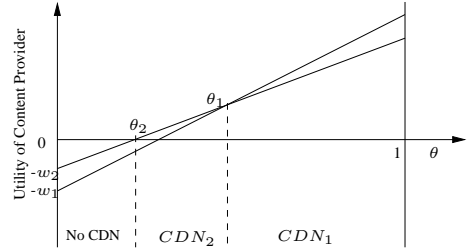}
\caption{Price setting where both CDNs have some revenue.}
\label{fig:feaprice}
\end{center}
\end{figure}

Now, $$\frac{dJ_1}{dw_1} = 0$$
$$\implies \left(1-\frac{w_1 - w_2}{\beta_2 -
\beta_1}\right)\times\Lambda - \frac{w_1\times\Lambda}{\beta_2 - \beta_1} =0$$
\begin{equation}
\label{eqn:cdncompp1}
\implies w_1 = \frac{\beta_2 -\beta_1 + w_2}{2}
\end{equation}

Now for $CDN_2$, 
$$\frac{dJ_2}{dw_2} = 0$$
\begin{eqnarray}
\implies &&\left(\frac{w_1 - w_2}{\beta_2 - \beta_1}-
\frac{w_2}{1-\beta_2}\right)\times\Lambda\nonumber\\
&& +  \left(\frac{-1}{\beta_2 -\beta_1} -
\frac{1}{1-\beta_2} \right)\times\Lambda \times w_2= 0\nonumber
\end{eqnarray}

\begin{equation}
\label{eqn:cdncompp2}
\implies w_2 = \frac{w_1\times(1-\beta_2)}{2(1-\beta_1)}
\end{equation}

Since the second derivatives are negative in both cases, the values we get are local maximum. So for a fix value of %%@
$w_2$, (\ref{eqn:cdncompp1}) gives the optimal value of $w_1$ for $CDN_1$; for a fix value of $w_1$, %%@
(\ref{eqn:cdncompp2}) gives the optimal value of $w_2$ for $CDN_2$. Now, if we solve (\ref{eqn:cdncompp1}) and %%@
(\ref{eqn:cdncompp2}), we get the equilibrium values $w_1^*$ and $w_2^*$ which are:
$$w_1^*=\frac{2(1-\beta_1)(\beta_2 -
\beta_1)}{4(1-\beta_1) - (1-\beta_2)},$$
$$w_2^*=\frac{(1-\beta_2)(\beta_2 - \beta_1)}{4(1-\beta_1)
- (1-\beta_2)}.$$

The	pair of prices $(w_1^*,w_2^*)$ gives the Nash Equilibrium, because if either of the CDN changes its price, then %%@
its revenue will go down.   

Now, since $\beta_2>\beta_1$, i.e., $(1-\beta_2) < (1
-\beta_1)$, we have,
$$w_1^* > \frac{2(1-\beta_2)(\beta_2 -
\beta_1)}{4(1-\beta_1) - (1-\beta_2)}$$
\begin{equation}
\label{eqn:twoprice}
\implies w_1^* > 2 w_2^* 
\end{equation}

Now, total number of content providers of $CDN_1$ is 
\begin{eqnarray*}
\Lambda_1^*&=& (1 - \theta_1^*)\Lambda  
= \left(1-\frac{w_1^* - w_2^*}{\beta_2
- \beta_1}\right)\Lambda \\
&=&\left(\frac{2(1-\beta_1)}{4(1-\beta_1) -
(1-\beta_2)}\right)\Lambda 
\end{eqnarray*} 
and the total number of content providers of $CDN_2$ is
\begin{eqnarray*}
\Lambda_2^*&=&(\theta_1^*-\theta_2^*)\Lambda
 = \left(\frac{w_1^* -
w_2^*}{\beta_2 - \beta_1} -\frac{w_2^*}{1-\beta_2} \right)\Lambda
\\
&=&\left(\frac{(1-\beta_1)}{4(1-\beta_1) -
(1-\beta_2)}\right)\Lambda 
\end{eqnarray*}
So we have,
\begin{equation}
\label{eqn:twousers}
\Lambda_1^*=2 \times \Lambda_2^*
\end{equation}
Now, equilibrium revenue of $CDN_1$ is  
$J_1^* = w_1^*\Lambda_1^*.$
Using (\ref{eqn:twoprice}) and (\ref{eqn:twousers}), we have
$J_1^*> 4 w_2^*\Lambda_1^*.$
So, $$J_1^* > 4 J_2^*.$$
\end{proof}

Note that, Theorem~\ref{thm:tworevenue} gives a conservative estimate which is always true
when $\beta_1>\beta_2$. The ratio of revenues increases significantly when the
difference between $\beta_1$ and $\beta_2$ increases.

\subsection{Competition among Three CDNs}

Let, three competitive CDNs, say, $CDN_1$, $CDN_2$, $CDN_3$,
have performance paratmeter $\beta_1$, $\beta_2$, $\beta_3$;
number of users $\Lambda_1$, $\Lambda_2$, $\Lambda_3$, and charge price
$w_1$, $w_2$, $w_3$, respectively.

\begin{theorem}
\label{thm:threerevenue}
If $\beta_1 <\beta_2 < \beta_3$, then in equilibrium,
$\Lambda_1^*>\Lambda_2^*>2 \Lambda_3^*$ and  $w_1^* > 1.5w_2^* > 3w_3^*$.
So in equilibrium, the revenue of $CDN_2$ is at least four times of that of $CDN_3$ and 
the revenue of $CDN_1$ is at least six times of that of $CDN_3$ i.e., 
$J_1^* > 1.5J_2^* > 6J_3^*.$
\end{theorem}
\begin{proof}
Similar techniques of the proof of Theorem~\ref{thm:tworevenue} can be used to proof this theorem. 
Here we provide a brief proof sketch by mentioning values for key terms. For complete proof of this theorem, %%@
see~\cite{Alam:2004}.

Equilibrium prices:
\begin{eqnarray*}
w_1^* &=& \Big[4(1-\beta_2)(\beta_2 - \beta_1)(\beta_3 -
\beta_1) -\\
& & (1-\beta_3)(\beta_2 - \beta_1)(\beta_2 -
\beta_1)\Big] \Big/\\ & &\Big[8(1-\beta_2)(\beta_3 - \beta_1) 
-2(1-\beta_3)(\beta_2 -\beta_1) -\\& & 2(1-\beta_2)(\beta_3 -
\beta_2)\Big],
\end{eqnarray*}
\begin{eqnarray*}
w_2^* &=& \Big[(1-\beta_2)(\beta_2 - \beta_1)(\beta_3 -
\beta_2)\Big] \Big/ \\ & & \Big[4(1-\beta_2)(\beta_3 - \beta_1) 
-(1-\beta_3)(\beta_2 -\beta_1)-\\ & & (1-\beta_2)(\beta_3 -
\beta_2)\Big]
\end{eqnarray*}
and
\begin{eqnarray*}
w_3^* &=& \Big[(1-\beta_3)(\beta_2 - \beta_1)(\beta_3 -
\beta_2)\Big] \Big/\\& & \Big[8(1-\beta_2)(\beta_3 - \beta_1) 
-2(1-\beta_3)(\beta_2 -\beta_1) -\\ & & 2(1-\beta_2)(\beta_3 -
\beta_2)\Big].
\end{eqnarray*}
Since $\beta_3>\beta_2$, i.e., $(1-\beta_3) < (1
-\beta_2)$, we have,
\begin{equation}
\label{eqn:p2p3}
w_2^* > 2 w_3^*
\end{equation}
and $(1-\beta_2)(\beta_2 - \beta_1)(\beta_3 - \beta_1) >
(1-\beta_3)(\beta_2 - \beta_1)(\beta_2 - \beta_1)$, so we
have,
\begin{equation}
\label{eqn:p1p2}
w_1^* > \frac{3}{2} w_2^*
\end{equation}

Number of content providers in equilibrium: 
\begin{eqnarray*}
\Lambda_1^* &=& \Big[2(1-\beta_2)(\beta_3 - \beta_1)\Lambda-\frac{1}{2}(1-\beta_3)(\beta_2-\beta_1)\Lambda\Big%%@
]\Big/\\ 
& & \Big[4(1-\beta_2)(\beta_3 - \beta_1)- (1-\beta_3)(\beta_2 -\beta_1) - \\
& & (1-\beta_2)(\beta_3 - \beta_2)\Big],
\end{eqnarray*}
\begin{eqnarray*}
\Lambda_2^* &=& \Big[(1-\beta_2)(\beta_3 - \beta_1)\Lambda + \frac{1}{2}(1-\beta_3)(\beta_2 - \beta_1)\Lambda \Big] %%@
\Big/ \\
& & \Big[4(1-\beta_2)(\beta_3 - \beta_1) -(1-\beta_3)(\beta_2 -\beta_1) - \\
& & (1-\beta_2)(\beta_3 - \beta_2)\Big]
\end{eqnarray*}
and 
\begin{eqnarray*}
\Lambda_3^* &=& \Big[\frac{1}{2}(\beta_2 - \beta_1)(\beta_3 -
\beta_2)\Lambda\Big] \Big / \Big[
4(1-\beta_2)(\beta_3 - \beta_1) 
- \\ & &(1-\beta_3)(\beta_2 -\beta_1) - (1-\beta_2)(\beta_3 -
\beta_2)\Big]. 
\end{eqnarray*}
Now, since $1 - \beta_2 > 1- \beta_3$ and $\beta_3-\beta_1
> \beta_2- \beta_1$, 
we have $(1 - \beta_2)(\beta_3-\beta_1) > (1-
\beta_3)(\beta_2- \beta_1),$ so, 
\begin{equation}
\label{eqn:w1w2}
\Lambda_1^* > \Lambda_2^* 
\end{equation}
Again, since $\beta_3-\beta_2 \leq 1 -\beta_2$ and $\beta_2 - \beta_1 < \beta_3 - \beta_1$, we have,
$(\beta_2 - \beta_1)(\beta_3-\beta_2) < (1-
\beta_2)(\beta_3- \beta_1),$ so,
\begin{equation}
\label{eqn:w2w3}
2\Lambda_3^* < \Lambda_2^* 
\end{equation}

From (\ref{eqn:w1w2}) and (\ref{eqn:w2w3}), we have $\Lambda_1^* 
> \Lambda_2^* >2\Lambda_3^*$ and
from (\ref{eqn:p2p3}) and (\ref{eqn:p1p2}), 
we have $w_1^* > 1.5w_2^* > 3 w_3^*$ in equilibrium. Since, $J_1^*=\Lambda_1^*w_1^*$,
$J_2^*=\Lambda_2^*w_2^*$ and $J_3^*=\Lambda_3^*w_3^*$, clearly, 
$J_1^* > 1.5J_2^* >6 J_3^*.$ 
\end{proof}

Although Theorem~\ref{thm:threerevenue} gives a modest lower bound, 
numerical results
of next section show that in most cases the performance advantage generates 
far higher revenue than 
this lower bound.

\subsection{Numerical Results}
Here we present some numerical results from MATLAB simulation for various values
of $\beta$s. First we show results for our first model of two CDNs and then we present
results for three CDNs. 

\begin{table}[h]
\begin{center}
\caption{Ratio of equilibrium revenues of two competitive CDNs for different values of $\beta$.}
\label{table:twocdn1}
\begin{tabular}{|c|c|c|}
\hline
$\beta_1$ & $\beta_2$ & $\frac{J_1^*}{J_2^*}$ \\ \hline
0.01 & 0.02 & 4.040816 \\
0.11 & 0.12 & 4.045455 \\
0.21 & 0.22 & 4.051282 \\
0.31 & 0.32 & 4.058824 \\
0.41 & 0.42 & 4.068966 \\
0.51 & 0.52 & 4.083333 \\
0.61 & 0.62 & 4.105263 \\
0.71 & 0.72 & 4.142857 \\
0.81 & 0.82 & 4.222222 \\
0.91 & 0.92 & 4.500000 \\
\hline
\end{tabular}
\end{center}
\end{table}

\begin{figure}
\centering
\includegraphics[height=2in]{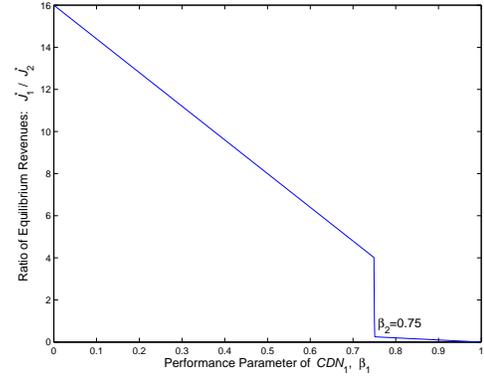}
\caption{Ratio of equilibrium revenues of $CDN_1$ and $CDN_2$ for different
values of $\beta_1$ when $\beta_2$ is fixed at 0.75.}
\label{fig:75twocdn}
\end{figure}

Results of Table~\ref{table:twocdn1} shows that when the difference of $\beta_1$ and $\beta_2$
is very small, equilibrium revenue of $CDN_1$ is almost always slightly higher than
four times of equilibrium revenue of $CDN_2$. Next, we keep $\beta_2$ fixed at 0.50 and check
the ratio of equilibrium revenues for different values of $\beta_1$ (see 
Fig.~\ref{fig:75twocdn}). It is clear that
the more the difference between the values of $\beta_1$ and $\beta_2$, the higher the 
ratio of equilibrium revenues.

\begin{table}[h]
\begin{center}
\caption{Ratios of equilibrium revenues of three competitive CDNs for different values of $\beta$.}
\label{table:threecdn1}
\begin{tabular}{|c|c|c|c|c|c|}
\hline
$\beta_1$ & $\beta_2$ & $\beta_3$ & $\frac{J_1^*}{J_2^*}$ & $\frac{J_2^*}{J_3^*}$&$\frac{J_1^*}{J_3^*}$\\ \hline
0.01 & 0.02 & 0.03 & 4.924346 & 988.082474 & 4865.659794 \\ 
0.11 & 0.12 & 0.13 & 4.927120 & 888.091954 & 4375.735632 \\ 
0.21 & 0.22 & 0.23 & 4.930608 & 788.103896 & 3885.831169 \\ 
0.31 & 0.32 & 0.33 & 4.935125 & 688.119403 & 3395.955224 \\ 
0.41 & 0.42 & 0.43 & 4.941206 & 588.140351 & 2906.122807 \\ 
0.51 & 0.52 & 0.53 & 4.949834 & 488.170213 & 2416.361702 \\ 
0.61 & 0.62 & 0.63 & 4.963033 & 388.216216 & 1926.729730 \\ 
0.71 & 0.72 & 0.73 & 4.985740 & 288.296296 & 1437.370370 \\ 
0.81 & 0.82 & 0.83 & 5.034020 & 188.470588 & 948.764706 \\ 
0.91 & 0.92 & 0.93 & 5.206731 & 89.142857 & 464.142857 \\ 
\hline
\end{tabular}
\end{center}
\end{table}

Although theoretically the lower bound we achieve for three competitive CDNs are
relatively modest compare to the lower bound of two CDNs, 
Table~\ref{table:threecdn1} shows that in the case of three competitive CDNs,
the ratio of equilibrium revenues for the top two CDNs are almost same as that for
two competitive CDNs. This result indicates that with the increase of number of competitors
the advantage of the CDN having best performance remains very significant. Another important
result is that in equilibrium the CDN having worst performance gets much smaller 
revenue compare to other CDNs even with very small difference in their values
of $\beta$.

\subsection{Discussions}
In the model analyzed,  it is found that 
the performance of a CDN is very critical in a competitive
market, i.e., a little performance advantage 
over the competitor may result in significant increase of revenue.
This results underscore the fact that during 
SLA (service level agreement), each CDN would like to announce the
best performance it could achieve in order to 
grab larger share of the market. 
However, if it can not achieve
that performance in real life, it has to pay 
huge penalty. This two facts imply the importance of designing a CDN with a
view to achieve that performance level as 
much as possible. Next section describes how this idea can be incorporated in the
request routing policy employed by a web-based CDN. 

\section{Routing Policies Based on Service Level Agreements}
\label{sec:routingpolicy}
Here, we concentrate on the design of a CDN from the perspective
 of  maximizing revenue.
We assume that in a typical SLA 
the CDN guarantees a minimum absolute (i.e.,  a fixed user 
latency), or relative latency (i.e., a fixed $\beta$) and charges a price for it.
 If the CDN can not achieve the specified $\beta$  for any request then it 
has to pay some penalty for that. The penalty is proportionate to the number 
of requests that can not be served within $\beta$.
It is assumed that there is no monetary incentive for the CDN to achieve $\beta$ better 
than what is specified in the SLA.
So maximizing revenue of a CDN actually depends on how many requests 
it can serve within a specified level of $\beta$.
The $\beta$ of a CDN is relevant to many design issues, such as
surrogate server allocation and placement, 
 request routing etc. 

For an existing CDN, little can be done about surrogate server allocation and
placement because the infrastructure can hardly be changed overnight. On the other hand, 
in the long term the existing infrastructure of other CDNs changes as well as new competitors
enter the market. It requires much more market oriented information to 
understand how the CDN react to competition in the long term. So we rather concentrate
on a model to determine request routing policy when the infrastructure (e.g., allocation and 
placement of surrogate servers) of a CDN is fixed. 

\subsection{Background}
The request routing problem is to decide the appropriate surrogate server for a 
particular client request in terms of certain metrics, e.g., 
surrogate server load (where surrogate server with the lowest load is chosen),
end-to-end latency (where surrogate server that offers the shortest response time
to the client is chosen), or distance (where surrogate server closest to the client
is chosen). Here, we look at the problem from a different perspective. 
We assume that $\beta$ is already set by the service level agreement (SLA) and what
we need is to find a request routing policy to select a surrogate server so that the number of requests 
that can be served within the required value of $\beta$ is maximized.

The entire request routing problem can be split into two parts: devising a redirection
 policy and selecting a redirection mechanism. A redirection policy defines how to 
 select a surrogate server in response to a given client request. It is basically an algorithm
 invoked when the client request is made. A redirection mechanism, in turn, is a mean of
 informing the client about this selection. Here, our focus is on the redirection policy.
 
 A redirection policy can be either adaptive or non-adaptive. The former considers 
 current system conditions while selecting a surrogate, whereas the latter does not.
 Adaptive redirection policies usually take higher selection time, but they have 
 lower transmission and processing time due to better selection. Several adaptive
(see ~\cite{Wang:2002}~\cite{Huffaker:2002}~\cite{Rodriguez:2000}~\cite%%@
{Andrews:2002}~\cite{Ardaiz:2001}
~\cite{Huffaker:2002}~\cite{Delgadillo:1999})
and nonadaptive policies 
(see  ~\cite{Pai:1998}~\cite{Rabinovich:1999}~\cite{Delgadillo:1999}) have been explored in the literature.
 
 A complex adaptive policy is used in Akamai~\cite{Dilley:2002}. It considers a few 
 additional metrics, like surrogate server load, the reliability of routes between 
 client and each of the surrogate servers, and bandwidth that is currently available
  to a surrogate server. Unfortunately, the actual policy is subject to trade secret
  and cannot be found in the published literature.
  
\subsection{Redirection Policy based on $\beta$}
In this section, we investigate about the redirection policy to maximize the monetary benefit of the 
CDN. We use a simplified model for the operation of a CDN and then analyze it to determine appropriate 
redirection policy for that model. First, we use Dynamic Programming (DP) to find the optimal policy.
However, any dynamic policy has large overhead in terms of online computation time. 
On the other hand, a static policy does not require any online computation, 
easier to implement and the overhead is low, if not zero.
For our model, we propose a static request redirection 
policy and prove that it is asymptotically optimal, i.e., optimal
for a system having very large arrival rate and service rate. Note that in the past static policy has been found %%@
asymptotically optimal in different contexts (e.g., congestion dependent network pricing~\cite{Paschalidis:2000}).

\subsubsection{The Model}
Here we introduce a model for the operation of a CDN. 
For simplicity, we assume that all requests are identical
(i.e., take same  resources such as  bandwidth, 
processor time etc., and the price and the penalty for every request are also same). 
We also assume that the origin server is far away and so, the time to  serve any 
request from the origin server
is assumed to be same (i.e., old user  latency is same for all end users). Then the required new user 
latency $\psi$ (=$\beta\times$old user latency) is same for all end users. 
Alternatively, this assumption is not required where the CDN guarantees a fixed user latency (i.e., $\psi$)
for the end users in the service level agreement with the content providers. For any request, if the CDN cannot serve %%@
it within $\psi$, then 
it has to pay a penalty.

\begin{figure}
\centering
\input{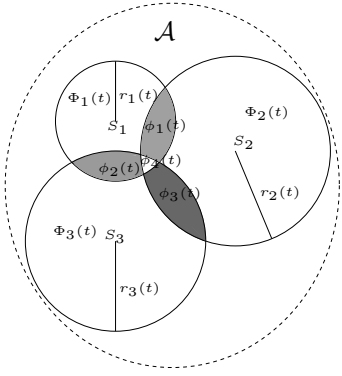}
\caption[An illustration of the term 'common area' and 'exclusive area' for a CDN having three surrogate servers.]
{An illustration of the term 'common area' and 'exclusive area' for a CDN having three surrogate servers. 
Dotted line represents the total area ${\cal A}$ of the network. End users are uniformly 
distributed over this area ${\cal A}$. At any time $t$, $\phi_1(t)$ is the common area
between server $S_1$ and $S_2$. Requests originated from this area can be served within the required value of 
$\beta$ (i.e. $\psi$)  by any of these two servers. Similarly, 
$\phi_2(t)$ is the common area between server $S_1$ and $S_3$, $\phi_3(t)$ is the common area
between server $S_2$ and $S_3$, $\phi_4(t)$ is the common area
among servers $S_1$, $S_2$ and $S_3$. At any time $t$, requests originating from area $\Phi_1(t)$ can
only be served by server $S_1$ and so $\Phi_1(t)$  is the  exclusive area of server $S_1$. Similarly, $\Phi_2(t)$
and $\Phi_3(t)$ are the exclusive areas of servers $S_2$ and $S_3$, respectively.}
\label{fig:commonarea}
\end{figure}

Let, there are total $m$ surrogate servers and the total area of the network
is ${\cal A}$ (see Fig.~\ref{fig:commonarea}). 
The end users are uniformly distributed over the whole network and 
the exponential request arrival rate in total network ${\cal A}$
is $\lambda$. So arrival rate per unit area is $\frac{\lambda}{\cal A}$.  Let, $\mu_1,\mu_2,...\mu_m$ be the %%@
exponential service rates of surrogate servers $1, 2,...,m$,
respectively (this service includes facing from the origin server, if needed). 
Let $w_1$ be the price charged by a CDN for serving a request within $\psi$ and ${w_1}'$ be 
the penalty for not serving it within the stipulated time.
It is also assumed that the servers have unlimited queuing capacity.

Let, $n_i(t)$ be the number of requests pending in surrogate server $S_i$ at time $t$. We will
be writing $\mathbf{N}(t)=(n_1(t), \cdots, n_m(t))$. We assume that the requests are processed in %%@
first-come-first-serve basis.  Then
on average the processing time for a request 
arriving at time $t$ in surrogate server $S_i$  
is $q_i(t)= \frac{n_i(t)+1}{\mu_i}$.

It is assumed that the time required for the end users to send a request to a surrogate
server is negligible and hence ignored. Now, a request that arrives 
at time $t$ can be served by surrogate server $S_i$ 
within a required value of $\beta$ if the transmission time from the 
end user and the surrogate server is less than or equal to $r_i(t) = \psi- q_i(t)$.
It is assumed that the transmission time is proportional to the distance and does not vary over time 
(i.e., we ignore any time varying network congestion). 
So any request originating at time $t$ within the radius $r_i(t)$ of server $S_i$ can be served 
by that server within the required value of $\beta$ (i.e., $\psi$). Say, this area, characterized by
the radius $r_i(t)$, is $A_i(t)$. Now if this area of 
two or more servers intersect at any time, then the requests which arrive from the 
common area can be served by any of these servers. If a request can be served by more than one surrogate servers
within the required value of $\beta$ (i.e. $\psi$),
then the policy to determine the appropriate server with an objective to maximize the revenue can be described as 
the routing policy we are looking for. If no server can serve the request within $\psi$, then it is served by the
origin server. So, we also ignore the availability issue of the CDN in our model.

\subsubsection{Analysis}

At first, we focus on some important features of the model that give some idea about the potential solutions.

\begin{lemma}
Under any policy the number of pending requests at any server $S_i$ is bounded by $\lceil \psi\mu_i-1 \rceil$.
\end{lemma}
\begin{proof}
Recall that the radius of any server $S_i$ is defined as $r_i(t)= \psi - \frac{n_i(t)+1}{\mu_i}$, where $n_i(t)$ is 
the number of requests at server $S_i$ at time $t$. So when the number of
requests at server $S_i$ is $\lceil \psi\mu_i-1 \rceil$, the radius $r_i \leq 0$. That means that when at any server %%@
$S_i$, the number of pending  
requests is $\lceil \psi\mu_i-1 \rceil$, no more requests can be served within the required value of $\psi$ and so %%@
the arrival rate is zero.
When the arrival rate is zero, the queue can not grow any more and so,
the maximum number of pending 
requests in any surrogate server $S_i$ is bounded by $\lceil \psi\mu_i -1 \rceil$.
\end{proof}

One important question for any routing policy is if any given request that can be served within $\psi$ should be %%@
served or not. It may be the case that
not serving the request may give the opportunity to serve more requests in the future. Following lemma shows that %%@
that is not the case.

\begin{lemma} 
\label{lem:allrequest}
If a request can be served by at least one surrogate server within the required
value of $\psi$, then on average serving the request returns more or at least equal revenue than not
serving it.
\end{lemma}
\begin{proof}
Suppose, there are two identical systems ${\cal S}$ and ${\cal S}'$ such that the number of surrogate servers,
their locations, service rates and the request arrivals everything is identical in both systems. Assume that system %%@
${\cal S}'$ use an optimal request routing policy and it denies
service to a certain request, say $i$-th request, even when it can be served by a server, say
server $S_i$. Now if we can show that another policy in system ${\cal S}$ that admits the $i$-th request
to  surrogate server $S_i$ and on average it can make at least as much revenue as system ${\cal S}'$, 
then we are done.
 
Suppose that $i$-th request is admitted to the server $S_i$ of system ${\cal S}$. Assume 
system ${\cal S}$ and ${\cal S}'$ takes exactly same decision for all requests except $i$-th request. Now
assume $j$-th request ($j>i$) is the first request after $i$-th request where system ${\cal S}$ can not
serve the request within the required value of $\psi$ but system ${\cal S}'$ can. So, system ${\cal S}$ can not
admit that request and lose revenue. Now, when $j$-th request arrives, in the worst case, 
server $S_i$ in system ${\cal S}$ has one more request to serve than the corresponding server in system
${\cal S}'$ and all other servers in both system are in identical state. So, in system ${\cal S}'$, 
server $S_i$ is the only server that can serve $j$-th request within the required value
of $\psi$. Now, when system ${\cal S}'$ accepts the $j$-th request, the revenue of both systems
are same. Since all requests are identical and both $i$-th and $j$-th requests
are served by server $S_i$ in system ${\cal S}$ and ${\cal S}'$, respectively, on average $i$-th request will depart
earlier in system ${\cal S}$ than $j$-th request in system ${\cal S}'$. So after that there would not be any request 
that can be served by system ${\cal S}'$, but not
by system ${\cal S}$, and on average the revenue of system ${\cal S}'$ is not 
greater than system ${\cal S}$. Note that, since on average $i$-th requests departs earlier than the arrival of 
$j$-th request, on average total revenue of system ${\cal S}$ will actually be higher than system ${\cal S}'$.  
\end{proof}

Following theorem defines a criteria to maximize the revenue.
\begin{theorem}
\label{thm:area}
The revenue is maximized if and only if the throughput of the system is maximized. The throughput of the system is %%@
maximized
if $$\lim_{T \rightarrow \infty} \frac{1}{T} E\left[\int_0^T 
\left(A_1(t)\cup A_2(t)\cup ... \cup A_m(t)\right) dt \right]$$ is maximized and all requests that can be served %%@
within $\beta$ (i.e., $\psi$)
are actually served by a surrogate server.
\end{theorem}
\begin{proof}
Since all requests have identical price and penalty, obviously maximizing throughput maximizes revenue and vice %%@
versa.

Now, at any time $t$ the number of requests that can be 
served within $\psi$ is $\lambda \frac{(A_1(t)\cup A_2(t)\cup ... \cup A_m(t))}{A}$ and the number of requests that %%@
can not
be served within $\psi$ is $\left(1- \lambda \frac{(A_1(t)\cup A_2(t)\cup ... \cup A_m(t))}{A}\right)$. 
Since from Lemma~\ref{lem:allrequest}, we know that not serving any request that can be served within $\psi$ does not %%@
increase overall revenue,
serving all the requests originating from the area $(A_1(t)\cup A_2(t)\cup ... \cup A_m(t))$ will maximize the %%@
throughput and revenue.
So net revenue from the requests originating
at time $t$ is 
\begin{eqnarray*}
\lefteqn{\lambda \frac{(A_1(t)\cup A_2(t)\cup ... \cup A_m(t))}{A} w_1 - }\\
& \left(1- \lambda \frac{(A_1(t)\cup A_2(t)\cup ... \cup A_m(t))}{A}\right){w_1}' \\
& =\lambda \frac{(A_1(t)\cup A_2(t)\cup ... \cup A_m(t))}{A} (w_1+{w_1}') - {w_1}'.
\end{eqnarray*}
So, the expected long term average revenue is given by
\begin{eqnarray*}
&\lim_{T \rightarrow \infty} \frac{\lambda}{T} E\Big[\int_0^T 
\Big(\frac{(A_1(t)\cup A_2(t)\cup ... \cup A_m(t))}{A}(w_1+{w_1}') -\\ 
& {w_1}' \Big) dt \Big].
\end{eqnarray*}
Since $\lambda, A, w_1, {w_1}'$ all are constants, above entity will be maximized if and only if the following entity %%@
is maximized
\begin{equation}
\label{eqn:networkpart_eqn1}
\lim_{T \rightarrow \infty} \frac{1}{T} E\left[\int_0^T 
\left(A_1(t)\cup A_2(t)\cup ... \cup A_m(t)\right) dt \right]
\end{equation}
(The above limit exists for any routing policy, because the state
$\mathbf{N}=0$, corresponding to an empty system, is recurrent.) 
\end{proof}

Since maximizing throughput maximizes revenue, from now on, we would try to find the policy that maximizes throughput %%@
and
would not use the terms $w_1$ and ${w_1}'$ anymore. 

\subsubsection{Optimal Dynamic Policy using Dynamic Programming}

Here we show how to obtain an optimal routing policy using 
{\it dynamic programming (DP)}.
Since the number of choices for each request is limited by 
the number of surrogate servers that can serve the request within $\beta$ (i.e. $\psi$), 
the set $\Pi$ of all possible policies ($\pi$) is compact. Again, in this finite state dynamic 
programming problem, all states communicate, i.e., for each pair of state $(\mathbf{N}, \mathbf{N'})$, there
exists a policy under which we can eventually reach $\mathbf{N'}$ starting from $\mathbf{N}$. So the state space is
irreducible. The Markov chain is also aperiodic. The expected holding time at any state is bounded by the service %%@
rate 
and the arrival rate at that state. The reward rate is bounded by the number of servers that are non-empty at that %%@
state. Under
these assumptions, the standard dynamic programming theory applies and asserts that there exists a policy which
is optimal.

The process $\mathbf{N}(t)$ is a continuous-time Markov chain. Since the total
transition out of any state is bounded by 
$\nu= \lambda +\sum_{i=1}^m \mu_i$, this Markov chain can be uniformized, 
leading  to a Bellman equation of the form
\begin{multline}
\label{eqn:bellman}
J^* +h(\mathbf{N}) = \max_{\pi \in \Pi} \Big[ \sum_{i=1}^m I(n_i(\mathbf{N}))+ \\
\sum_{i=1}^m 
\frac{\lambda_i^{\pi}(\mathbf{N})}{\nu} h(\mathbf{N}+e_i) +  \sum_{i=1}^m \frac{\mu_i I(n_i(\mathbf{N}))}{\nu} %%@
h(\mathbf{N}-e_i)\\
+\left(1-\sum_{i=1}^m \frac{\lambda_i^{\pi}(\mathbf{N})}{\nu} - \sum_{i=1}^m \frac{\mu_i %%@
I(n_i(\mathbf{N}))}{\nu}\right) h(\mathbf{N})\Big]
\end{multline}

Here, $\lambda_i^{\pi}(\mathbf{N})$ is the request arrival rate at server $S_i$ at system state $\mathbf{N}$ under %%@
policy $\pi$,
and to indicate the non-emptiness of a server we use $I(n_i(\mathbf{N}))$, where $n_i(\mathbf{N})$ denotes the number %%@
of requests
at server $S_i$ when the system is at state $\mathbf{N}$. We define $I(n_i(\mathbf{N}))$ as follows:
\begin{equation}
I(n_i(\mathbf{N}))=
\begin{cases}
1 & \text{if } n_i(\mathbf{N})>0\\
0 & \text{otherwise.} 
\end{cases}
\nonumber
\end{equation}

We impose the condition $h(0)=0$, in which case Bellman's equation has a unique solution 
[in the unknowns $J^*$ and $h(\cdot)$].
Once Bellman's equation is solved, an optimal policy is readily 
obtained by choosing the policy at each 
state $\mathbf{N}$ that maximizes the right-hand side in (\ref{eqn:bellman}).
The solution to Bellman's equation has the following interpretation: the scalar $J^*$ 
is the optimal expected
reward per unit time, and $h(\mathbf{N})$ is the relative reward in state $\mathbf{N}$.
In particular, consider an optimal policy that attains the maximum in (\ref{eqn:bellman})
for every state $\mathbf{N}$. If we follow this policy starting from $\mathbf{N}'$
or state $\mathbf{N}$, the expectation of the difference in total rewards (over the
infinite horizon) is equal to ${(h(\mathbf{N}')-h(\mathbf{N}))}/{\nu}$.

The solution to Bellman's equation and resulting optimal policy can be computed 
using classical DP algorithms. However, the computational complexity
increases with the size of state space, which is exponential in the
number of surrogate servers $m$. For this reason, exact solution using DP is 
only feasible only when the number of surrogate servers is quite small.

\subsection{Static Redirection Policy}
\label{sec:staticpolicy}
Here we focus on finding a static policy that is asymptotically optimal and at the same time the overhead is low (no %%@
runtime computation to select the server).

\noindent
{\it Definition:} We say that a routing policy is {\it static} if a particular surrogate server is always responsible %%@
to serve  
requests originating from a particular end user independent of the state of the system. 

The key idea of the proposed policy is distributing the common area among the participating servers in such a way %%@
that 
the total throughput is maximized as well as the load is balanced among the surrogate servers as much as possible. We %%@
calculate the proportions of each
common area to be assigned to a participating server when the system is at empty state. 

\subsubsection{Optimization Problem and Proposed static policy, $\pi^s$}
Suppose, when the system is at empty state the total number of common areas is $Z$, at empty state total arrival rate %%@
at common area $z$ is $\phi_z=\phi_z(0)$, 
at empty state total arrival rate at exclusive area 
(i.e., not common with other servers) of 
server $S_i$ is $\Phi_i = \Phi_i(0)$ (see Fig.~\ref{fig:commonarea}). Say, our policy assigns $P_{iz}$ portion of %%@
common area $z$ to server $S_i$. 
Let, the number of participating servers of common area $z$ is $z_a$ and they are denoted as $i_{z_1}...i_{z_a}$. %%@
Similarly, server $S_i$ participates in 
total $i_b$ common areas denoted as $z_{i_1}..z_{i_b}$.
Now, at empty state the total arrival rate for server $S_i$ is $\lambda_i(0)=\Phi_i + %%@
\sum_{z=z_{i_1}}^{z_{i_b}}P_{iz}\phi_z$.  

Then the optimization problem is
$$\max \sum_{i=1}^m I_i \left(\Phi_i + \sum_{z=z_{i_1}}^{z_{i_b}}P_{iz}\phi_z\right)$$
subject to 
\begin{eqnarray}
\label{eqn:policy}
&0 \leq I_i \leq 1 \;\; \forall i, \nonumber \\
&0 \leq I_i \left(\Phi_i + \sum_{z=z_{i_1}}^{z_{i_b}}P_{iz}\phi_z\right) \leq \mu_i \;\; \forall i, \nonumber\\
&0 \leq P_{iz}\leq 1 \;\;\;\;\forall i\;\forall z\;,\nonumber\\
&0 \leq \sum_{i=i_{z_1}}^{i_{z_b}} P_{iz} \leq 1 \;\;\;\;\forall z
\end{eqnarray}
Note that,  in the optimal solution of the above problem, 
if $\lambda_i> \mu_i$, then the value of $I_i$ will be set in such a way that $I_i\lambda_i= \mu_i$,
else if $\lambda_i \leq \mu_i$,  the value of $I_i$ will be 1. 
Once we have solve above problem and have the values of $P_{iz}$'s then we divide common area
$z$ among the servers such that each server $S_i$ gets $P_{iz}$ portion of the common area. 
So server $S_i$ will serve $P_{iz}$ portions of the requests that
originates from common area $z$. Say, this proposed policy is $\pi^s$.

The optimization problem ensures that 
$$\sum_{i=1}^m \min(\lambda_i(0), \mu_i) = \max_{\forall \pi \in \Pi} \sum_{i=1}^m \min(\lambda_i^{\pi}(0), \mu_i)$$ %%@
where
$\lambda_i(0)$ is the request arrival rate at server $S_i$ under our policy $\pi^s$ when the system is at empty %%@
state; $\Pi$ is the set
of all possible policies of distributing the common areas among the servers and $\lambda_i^{\pi}(0)$ is the request %%@
arrival rate at server $S_i$ under 
policy $\pi$ when the system is at empty state. We use this information to prove that policy $\pi^s$ is %%@
asymptotically optimal.

Finally, in the asymptotic analysis of next section, it is shown that any solution of the optimization problem %%@
(\ref{eqn:policy}) is sufficient
to be optimal for a system where the arrival and service rate is close to infinity. However, in most cases, the %%@
optimization problem (\ref{eqn:policy})
has multiple solutions. Now the optimization problem (\ref{eqn:policy1}) can be used to pick the best solution among %%@
those solutions so that
the system performs well even with small arrival and service rate.

If $\gamma$ denotes the 
optimal value of $\sum_{i=1}^m I_i \left(\Phi_i+\sum_{z=z_{i_1}}^{z_{i_b}}P_{iz}\phi_z\right)$, then the  %%@
optimization problem is
$$\min  \sum_{z=1}^{Z} \sigma(\lambda_{z_i}/\mu_{z_i})$$
subject to
\begin{eqnarray} 
\label{eqn:policy1}
&\sum_{i=1}^m I_i \left(\Phi_i + \sum_{z=z_{i_1}}^{z_{i_b}}P_{iz}\phi_z\right)=\gamma, \nonumber\\
&0 \leq I_i \leq 1 \;\; \forall i, \nonumber\\
&0 \leq I_i \left(\Phi_i + \sum_{z=z_{i_1}}^{z_{i_b}}P_{iz}\phi_z\right) \leq \mu_i \;\; \forall i, \nonumber\\
&0 \leq P_{iz}\leq 1 \;\;\;\;\forall i\;\forall z\;, \nonumber\\ 
&0 \leq \sum_{i=i_{z_1}}^{i_{z_b}} P_{iz} \leq 1 \;\;\;\;\forall z
\end{eqnarray}

This way of solving multiobjective problems sequentially is known as the lexicographical
method ~\cite{Fishburn:1974}~\cite{Szidar:1986}.
Note that, $\sigma(\lambda_{z_i}/\mu_{z_i})$ represents the variance of load (i.e., arrival rate/ service rate)
of the participating servers of common area $z$.

\subsubsection{Asymptotic Analysis}
Since under our proposed static policy $\pi^s$ the arrival rate in each surrogate server
does not depend on the states of other servers, the analysis can be carried out for
each individual server independently. At first, we analyze the case of a single server CDN
and then extend it for a multiple server CDN to show that our proposed policy is asymptotically
optimal i.e., ratio of the throughput of the proposed policy and the upper bound of the throughput 
under any policy converges to 1 when the arrival rate and the service rate of the system goes to infinity.

\noindent
{\em Throughput for a Single Server CDN}

\begin{figure}
\input{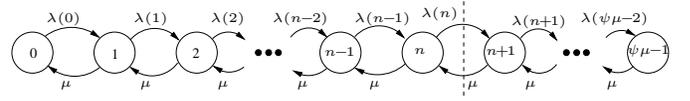}
\caption{State diagram for a single server CDN.}
\label{fig:singlestate}
\end{figure}

Consider a CDN consisting of one server which has exponential service rate $\mu$ and the exponential
arrival rate at state $n$  is $\lambda(n)=\frac{\lambda}{{\cal A}} \pi \left(\psi-\frac{n+1}{\mu} \right)^2,$
where $\frac{\lambda}{{\cal A}}$ is the state independent arrival rate per unit area and $\left(\psi-\frac{n+1}{\mu} %%@
\right)$ 
denotes the radius of the circular area covered by the server when there are
$n$ requests in the server. So this system can be viewed as a special $M/M/1$ queue where exponential arrival rate %%@
varies with the
change of system state. State diagram for the system is shown in Fig.~\ref{fig:singlestate}. 

We are interested in finding the throughput of such a CDN. Following lemma gives the exact value of the throughput.
\begin{lemma}
\label{lem:singlecdnthroughput}
The throughput of the single server CDN is 
$$J_1=\mu\left(1- \frac{1}{1+ \sum_{n=1}^{\lceil \psi\mu-1 \rceil} \prod_{l=0}^{n-1} \frac{\lambda(l)}{\mu}}\right)$$
\end{lemma}
\begin{proof}
Let $p_n$ be the probability that the system is at state $n$. Since the total number of states is finite (i.e. %%@
$n_{max} = \lceil \psi\mu-1 \rceil $), the system is stable. 
So over a long period of time, the number of transitions from any state $n$ to state $n+1$ equals the number of %%@
transitions from state $n+1$ to state $n$.
Thus we obtain the balance equations
$$p_n \lambda(n) = p_{n+1} \mu \;\;\; \text{for each } (\lceil \psi\mu-1 \rceil) > n \geq 0.$$
So, we have, $$p_1= \frac{\lambda(0)}{\mu} p_0,$$
$$p_2= \frac{\lambda(1)}{\mu} p_1= p_0 \frac{\lambda(1)}{\mu} \frac{\lambda(0)}{\mu},$$
and in general,
$$p_n= \frac{\lambda(1)}{\mu} p_{n-1}= p_0\prod_{l=0}^{n-1} \frac{\lambda(l)}{\mu}.$$

Since summation of the probabilities of all states is 1, we have
$$\sum_{n=0}^{\lceil \psi\mu-1 \rceil} p_n =1.$$
Putting the values of $p_n$, we have
$$p_0 + \sum_{n=1}^{\lceil \psi\mu-1 \rceil} p_0\prod_{l=0}^{n-1} \frac{\lambda(l)}{\mu} =1.$$

Solving for $p_0$, we get
$$p_0 = \frac{1}{1+ \sum_{n=1}^{\lceil \psi\mu-1 \rceil} \prod_{l=0}^{n-1} \frac{\lambda(l)}{\mu}}.$$

So the throughput is $$\mu(1-p_0)=\mu\left(1- \frac{1}{1+ \sum_{n=1}^{\lceil \psi\mu-1 \rceil} \prod_{l=0}^{n-1} %%@
\frac{\lambda(l)}{\mu}}\right).$$
\end{proof}

Now we use this result to do asymptotic analysis of the throughput of a single server CDN.

\noindent
{\em Asymptotic Analysis for a Single Server CDN}

For asymptotic analysis, we scale the system through a proportional increase in arrival rate and service rate. More %%@
specifically,
let $c \geq1$ be a scaling factor. The scaled system has arrival rate $\lambda^c=c\lambda$ and the service rate %%@
$\mu^c = c\mu$.
Note that other parameters, total network area ${\cal A}$ and maximum allowed user latency $\psi$ are held fixed. 
We will use a superscript $c$ to denote various quantities of interest for
the scaled system. 

At first, we would like to find an upper bound for the throughput for the single server.
\begin{lemma}
The upper bound for the throughput of a single server CDN is $$J_{ub}^c= \min(\lambda^c(0), \mu^c),$$ 
where $\lambda^c(n)$ denote the arrival rate when number of requests at the server is $n$ and 
$\mu^c$ is the service rate.
\end{lemma}
\begin{proof}
Since for all $n$, $\lambda^c(0) \geq \lambda^c(n)$, maximum arrival rate is $\lambda^c(0)$. Again, the server can %%@
never serve at a rate faster than $\mu^c$. So clearly the
throughput is bounded by $\min(\lambda^c(0), \mu^c).$
\end{proof}

Following lemma shows that the actual throughput of a very large system converges to this upper bound. It is easily %%@
seen that the upper bound $J_{ub}^c$ and actual 
throughput $J_1^c$ will increase roughly linearly with $c$, and for this reason, a meaningful comparison should first %%@
divide such quantities by $c$, as in the result
that follows.
\begin{lemma}
\label{lem:asympss}
For the single server CDN, $$\lim_{c \rightarrow \infty} \frac{1}{c}J_1^c =\lim_{c \rightarrow \infty} %%@
\frac{1}{c}J_{ub}^c,$$ where $J_1^c$ is the actual throughput of the scaled system and
$J_{ub}^c=\min(\lambda^c(0), \mu^c).$
\end{lemma}
\begin{proof}
Fix some $\epsilon>0$ and let us consider a system with arrival rate $\lambda_{\epsilon}^c(0)=\lambda^c(0)+\epsilon$ %%@
and other parameters $\mu^c, {\cal A}, \psi$ are kept same. Let 
$J_{1\epsilon}^c$ be the resulting actual throughput. Clearly, for every $n$, $\lambda_{\epsilon}^c(n) \geq %%@
\lambda^c(n).$
Now the probability of the empty state for this system is 
\begin{equation}
\label{eqn:proempty1}
p_{0\epsilon}^c = \frac{1}{1+ \sum_{n=1}^{c\mu \psi-1} \prod_{l=0}^{n-1} \frac{\lambda_{\epsilon}^c(l)}{\mu^c}}
\end{equation}
Since
$$\frac{\frac{\lambda^c(n)}{\mu^c}}{\frac{\lambda^c(0)}{\mu^c}}= \frac{\lambda^c(n)}{\lambda^c(0)}=\frac{(\psi - %%@
\frac{n+1}{c\mu})^2}{(\psi - \frac{1}{c\mu})^2}=\left(1- \frac{n}{c\mu \psi -1}\right)^2,$$
we have, 
$$\frac{\lambda_{\epsilon}^c(n)}{\mu^c} \geq \frac{\lambda^c(n)}{\mu^c} = \left(1- \frac{n}{c\mu \psi -1}\right)^2 %%@
\frac{\lambda^c(0)}{\mu^c}$$
So, for any finite constant ${\cal C}$,
$\lim_{c \rightarrow \infty} \frac{\lambda_{\epsilon}^c({\cal C})}{\mu^c} \geq \lim_{c \rightarrow \infty} %%@
\frac{\lambda^c(0)}{\mu^c}$
and then from (\ref{eqn:proempty1}), we have,
\begin{eqnarray}
\label{eqn:asympsingle}
p_{0\epsilon}^c &\leq& \frac{1}{1+ \sum_{n=1}^{\cal C} \left({\frac{\lambda^c(0)}{\mu^c}}\right)^n + \sum_{n={\cal %%@
C}+1}^{c\mu \psi-1} \prod_{l=0}^{n-1} \frac{\lambda_{\epsilon}^c(l)}{\mu^c}}\nonumber\\
&\leq& \frac{1}{1+ \sum_{n=1}^{\cal{C}} \left({\frac{\lambda^c(0)}{\mu^c}}\right)^n}
\end{eqnarray}

Now, the arrival rate at empty state of the scaled system can be greater than, equal to or less than the service %%@
rate.  We prove the lemma separately for those cases.

\noindent
{\em Case 1: $\lambda^c(0) < \mu^c$}

From (\ref{eqn:asympsingle}) we have,
\begin{eqnarray}
\lim_{c\rightarrow \infty} p_{0\epsilon}^c &\leq& \frac{1}{1+ \frac{\frac{\lambda^c(0)}{\mu^c}\left(1 - %%@
\left(\frac{\lambda^c(0)}{\mu^c}\right)^{\cal{C}}\right)}{1-\frac{\lambda^c(0)}{\mu^c}}}\nonumber\\
&=& \frac{1-\frac{\lambda^c(0)}{\mu^c}}{1 -\left(\frac{\lambda^c(0)}{\mu^c}\right)^{{\cal{C}}+1}}.\nonumber
\end{eqnarray}
When $c\rightarrow\infty$, we are free to choose an arbitrarily large value for $\cal{C}$, then the value of 
$\left(\frac{1-\frac{\lambda^c(0)}{\mu^c}}{1 -\left(\frac{\lambda^c(0)}{\mu^c}\right)^{{\cal{C}}+1}}\right)$ will %%@
converges to $\left(1-\frac{\lambda^c(0)}{\mu^c}\right)$ and
then we have, 
$$\lim_{c\rightarrow \infty} p_{0\epsilon}^c \leq \left(1-\frac{\lambda^c(0)}{\mu^c}\right).$$ 

So 
\begin{equation}
\label{eqn:singlethrou1}
\lim_{c \rightarrow \infty} \frac{1}{c} J_{1\epsilon}^c=\lim_{c \rightarrow \infty} \frac{1}{c} %%@
\mu^c(1-p_{0\epsilon}^c) \geq \lim_{c \rightarrow \infty} \frac{1}{c} \lambda^c(0)
\end{equation}

Again, since for all $n$, $\frac{\lambda_{\epsilon}^c(0)}{\mu^c} \geq \frac{\lambda_{\epsilon}^c(n)}{\mu^c}$, we have
$\lim_{c\rightarrow \infty} p_{0\epsilon}^c \geq \frac{1}{1+ \sum_{n=1}^{\infty} %%@
\left({\frac{\lambda_{\epsilon}^c(0)}{\mu^c}}\right)^n}$, which implies, 
$$\lim_{c\rightarrow \infty} p_0^c \geq \left(1-\frac{\lambda_{\epsilon}^c(0)}{\mu^c}\right).$$ 

So 
\begin{equation}
\label{eqn:singlethrou2}
\lim_{c \rightarrow \infty} \frac{1}{c} J_{1\epsilon}^c=\lim_{c \rightarrow \infty} \frac{1}{c} %%@
\mu^c(1-p_{0\epsilon}^c) \leq \lim_{c \rightarrow \infty} \frac{1}{c} \lambda_{\epsilon}^c(0)
\end{equation}

This is true for any positive $\epsilon$. We now let $\epsilon$ go to zero in which case $\lim_{c \rightarrow \infty} %%@
\frac{1}{c} \lambda_{\epsilon}^c(0)$ tends to 
$\lim_{c \rightarrow \infty} \frac{1}{c} \lambda^c(0)$, and from (\ref{eqn:singlethrou1}) and %%@
(\ref{eqn:singlethrou2}), we have
$$\lim_{c \rightarrow \infty} \frac{1}{c} J_{1}^c =\lim_{c \rightarrow \infty} \frac{1}{c} \lambda^c(0) = \lim_{c %%@
\rightarrow \infty} \frac{1}{c} J_{ub}^c.$$

\noindent
{\em Case 2: $\lambda^c(0) \geq \mu^c$}

When $\frac{\lambda^c(0)}{\mu^c} \geq 1$ and the value for ${\cal C}$ is arbitrarily large,  $\left(\frac{1}{1+ %%@
\sum_{n=1}^{\cal{C}} \left({\frac{\lambda^c(0)}{\mu^c}}\right)^n}\right)$
converges to zero. Since
when $c\rightarrow\infty$, we are free to choose an arbitrarily large value for ${\cal C}$ and 
probability can not be negative, 
from (\ref{eqn:asympsingle}) we have, $$\lim_{c\rightarrow \infty} p_{0\epsilon}^c=0.$$ 

This is true for any positive $\epsilon$. We now let $\epsilon$ go to zero in which case $\lim_{c \rightarrow \infty} %%@
\frac{1}{c} p_{0\epsilon}^c$ tends to 
$\lim_{c \rightarrow \infty} \frac{1}{c} p_0^c$, 
and we have,
$$\lim_{c \rightarrow \infty} \frac{1}{c} J_{1}^c =\lim_{c \rightarrow \infty} \frac{1}{c} \mu^c = \lim_{c %%@
\rightarrow \infty} \frac{1}{c} J_{ub}^c.$$
\end{proof}

Although it seems that $c\rightarrow \infty$ is not practically feasible, a very good approximation is achieved for %%@
fairly small values of $c$ (see 
Table~\ref{table:singlebiglam},~\ref{table:single},~\ref{table:sin%%@
glesmalllam}).

\noindent
{\em Asymptotic Analysis for Multiple Server CDN under Static Policy, $\pi^s$}

For the multiple server case we scale the system in the same way like the asymptotic analysis for single server CDN. %%@
Let, ${\cal J}_s^c$ be the actual 
throughput of our static policy $\pi^s$ in the scaled system and ${\cal J}_{ub}^c$ be the upper bound of the %%@
throughput under any policy.

To assess the degree of suboptimality of our policy, we develop an upper bound of throughput for the multiple server %%@
CDN
under any policy.

\begin{lemma}
\label{lem:throughmul}
For multiple server CDN, upper bound of throughput is 
$${\cal J}_{ub}^c = \sum_{i=1}^m \min\left( \lambda_i^c(0), \mu_i^c \right)$$
where 
$\lambda_i^c(0)$ is the maximum available arrival rate for surrogate server $S_i$ at empty state after distributing
the arrival rates in the common areas among the servers in such a way that $\sum_{i=1}^m \min\left( \lambda_i^c(0), %%@
\mu_i^c \right)$ is maximized.
\end{lemma}
\begin{proof}
Since any server can not serve at a rate higher than its service rate, 
the upper bound of the throughput is actually the minimum of the service rate and the arrival rate. 
The area covered by a server is maximum when it is idle (i.e., no outstanding request). So, when the system is at %%@
empty state, total
arrival rate is maximum. Now distributing the arrivals in the common area in such a way that maximizes the term 
$\sum_{i=1}^m \min\left( \lambda_i^c(0), \mu_i^c \right)$, where $\lambda_i^c(0)$ is the total arrival rate at server %%@
$S_i$ after distribution, ensures that
each server has maximum available arrival rate. Hence, the upper bound is determined by summation of the minimum of %%@
this arrival rate and the service rate
of each server. 
\end{proof}

Following theorem shows that for very large system our policy actually achieve this upper bound.
\begin{theorem}
If ${\cal J}_s^c$ is the actual throughput of policy  $\pi^s$ and ${\cal J}_{ub}^c$ is the upper bound of the %%@
throughput under any policy, then
$$\lim_{c \rightarrow \infty} \frac{1}{c}{\cal J}_s^c =\lim_{c \rightarrow \infty} \frac{1}{c}{\cal J}_{ub}^c.$$
\end{theorem}

\begin{proof}
Since under policy $\pi^s$, each user is mapped to one and only surrogate server, the system will behave like $m$ %%@
independent single server CDNs. So we can carry out the analysis
for each server independently without taking into account the state of other servers. Let ${\cal J}_i^c$ be the %%@
throughput of server $S_i$. Then we have, 
$${\cal J}_s^c=\sum_{i=1}^m {\cal J}_i^c.$$

In multiple server case, for any server $S_i$, the arrival rate may be less than the arrival rate in the total %%@
circular area encompassed by its radius due to sharing of
common areas with other servers. Suppose, when the server $S_i$ is at empty state the size
of the area inside its circle that is given to other servers is $a_{i,0}^c$ and when the server is at state $n$ the %%@
size of the area inside its circle that is given to other servers is $a_{i,n}^c$.

Fix some $\epsilon > 0$ and let us consider a system where the arrival rate in the whole network ${\cal A}$ is %%@
$\lambda_{\epsilon}^c = \lambda^c +\epsilon$ and other parameters
$\mu_i^c, {\cal A}, \psi$ are same as those of the original system. Due to uniform distribution of end users
in the whole network, at empty state arrival rate for any server would be $\lambda_{i\epsilon}^c(0)=\lambda_{i}^c(0) %%@
+\epsilon_i$ where $\epsilon_i \leq \epsilon$. Clearly for every $n$,
$\lambda_{i\epsilon}^c(n) \geq \lambda_{i}^c(n)$.

Since,
$$\frac{\frac{\lambda_i^c(n)}{\mu_i^c}}{\frac{\lambda_i^c(0)}{\mu_%%@
i^c}}= \frac{\lambda_i^c(n)}{\lambda_i^c(0)}=\frac{(\psi - \frac{n+1}{c\mu})^2 - a_{i,n}^c}{(\psi - %%@
\frac{1}{c\mu_i})^2 -a_{i,0}^c},$$
we have, 
$$\frac{\lambda_{i\epsilon}^c(n)}{\mu_i^c} \geq \frac{\lambda_i^c(n)}{\mu_i^c} = \left(\frac{(\psi - %%@
\frac{n+1}{c\mu})^2 - a_{i,n}^c}{(\psi - \frac{1}{c\mu})^2 -a_{i,0}^c}\right) \frac{\lambda^c(0)}{\mu^c}$$

Clearly $a_{i,0}^c \geq a_{i,n}^c \geq 0$ and $(\psi - \frac{1}{c\mu_i})^2- (\psi - \frac{n+1}{c\mu_i})^2 \geq %%@
\frac{(a_{i,0}^c-a_{i,n}^c )}{\pi}$. So when $c\mu_i \gg n$, $(\psi - \frac{1}{c\mu_i})^2 \approx
 (\psi - \frac{n+1}{c\mu_i})^2$ and $a_{i,0}^c \approx a_{i,n}^c$. So, for any finite constant ${\cal C},$ 
$\lim_{c \rightarrow \infty} \frac{(\psi - \frac{{\cal C}+1}{c\mu})^2 - a_{i,{\cal C}}^c}{(\psi - \frac{1}{c\mu})^2 %%@
-a_{i,0}^c}$ converges to 1.
So,  for any finite constant ${\cal C}$,
$$\lim_{c \rightarrow \infty} \frac{\lambda_{i\epsilon}^c(n)}{\mu_i^c} \geq \frac{\lambda^c(0)}{\mu^c}.$$

Now using exactly the same technique of Lemma~\ref{lem:asympss}, we have,
$$\lim_{c \rightarrow \infty} \frac{1}{c} {\cal J}_i^c = \lim_{c \rightarrow \infty} \frac{1}{c} %%@
\min\left({\lambda_i^c(0), \mu_i}\right).$$

Then,
$$\lim_{c \rightarrow \infty} \frac{1}{c} {\cal J}_s^c = \lim_{c \rightarrow \infty} \frac{1}{c} \sum_{i=1}^m {\cal %%@
J}_i^c= \lim_{c \rightarrow \infty} \frac{1}{c} \sum_{i=1}^m  \min\left({\lambda_i^c(0), \mu_i}\right)$$

But, by Lemma~\ref{lem:throughmul}, we have
$${\cal J}_{ub}^c = \sum_{i=1}^m \min\left( \lambda_i^c(0), \mu_i^c \right)$$
So, we have,
$$\lim_{c \rightarrow \infty} \frac{1}{c} {\cal J}_s^c = \lim_{c \rightarrow \infty} \frac{1}{c} {\cal J}_{ub}^c.$$
\end{proof}

The result of above theorem shows that our static policy $\pi^s$ has asymptotically optimal throughput and thus %%@
generate maximum revenue for very large system.

\subsection{Numerical  Results}

In this section, we present some numerical results from MATLAB simulation that show that actual throughput converges %%@
to the upper bound of the throughput when the system is large. 
We check the throughput for all three possible cases where arrival rate is greater than, equal to and less than the %%@
service rate for a single server case.
% empty state arrival rate higher than service rate
\begin{table}[h]
\begin{center}
\caption{Single server CDN with $\lambda(0)>\mu$; $\psi=1000$, $\lambda(0)=1.05$, $\mu=1$.}
\label{table:singlebiglam}
\begin{tabular}{|c|c|c|c|c|}
\hline
$\text{Scaling factor,} c$ & Throughput & Upper Bound & $\frac{\text{Throughput }}{\text{Upper bound}}$\\ \hline
1 & 0.988756 & 1.000000 & 0.988756 \\
2 & 1.991859 & 2.000000 & 0.995930 \\
3 & 2.994649 & 3.000000 & 0.998216 \\
4 & 3.996626 & 4.000000 & 0.999156 \\
5 & 4.997924 & 5.000000 & 0.999585 \\
6 & 5.998744 & 6.000000 & 0.999791 \\
7 & 6.999250 & 7.000000 & 0.999893 \\
8 & 7.999556 & 8.000000 & 0.999944 \\
9 & 8.999739 & 9.000000 & 0.999971 \\
10 & 9.999848 & 10.000000 & 0.999985 \\
11 & 10.999911 & 11.000000 & 0.999992 \\
12 & 11.999949 & 12.000000 & 0.999996 \\
13 & 12.999970 & 13.000000 & 0.999998 \\
14 & 13.999983 & 14.000000 & 0.999999 \\
15 & 14.999990 & 15.000000 & 0.999999 \\
16 & 15.999994 & 16.000000 & 1.000000 \\
\hline
\end{tabular}
\end{center}
\end{table}

\begin{table}[h]
\begin{center}
\caption{Single server CDN with $\lambda(0)=\mu$; $\psi=1000$, $\lambda(0)=1$ and $\mu=1$.}
\label{table:single}
\begin{tabular}{|c|c|c|c|}
\hline
$\text{Scaling factor,} c$& Throughput & Upper Bound & $\frac{\text{Throughput }}{\text{Upper bound}}$\\ \hline
1 & 0.9653 & 1 & 0.964094 \\
2 & 1.9506  &    2          & 0.974659\\
5 &  4.9213  &    5         &  0.983999 \\ 
10  &  9.8882    &  10         & 0.988695\\
20 &        19.8415   &  20         & 0.992010  \\
50 &        49.7487  &   50        &  0.994949\\
100   &        99.6442   &  100       &  0.996430\\
200    &       199.4964   & 200       &  0.997476\\
2000  &     1.9986e+03  & 2000     &   0.999311\\
20000 &    1.9995e+04   & 20000   &    0.999748 \\ 
200000  &    1.9998e+05  & 200000  &    0.999920\\
2000000 &   1.9999e+06  & 2000000   &  0.999975\\
\hline
\end{tabular}
\end{center}
\end{table}

\begin{table}[h]
\begin{center}
\caption{Single server CDN with $\lambda(0)<\mu$; $\psi=1000$, $\lambda(0)=0.8$ and $\mu=1$.}
\label{table:singlesmalllam}
\begin{tabular}{|c|c|c|c|}
\hline
$\text{Scaling factor,} c$& Throughput & Upper Bound & $\frac{\text{Throughput}}{\text{Upper bound}}$\\ \hline
1 & 0.794054 & 0.800000 & 0.992567 \\
100& 79.993605 & 80.000000 & 0.999920 \\
200& 159.993603 & 160.000000 & 0.999960 \\
300 & 239.993602 & 240.000000 & 0.999973 \\
400& 319.993601 & 320.000000 & 0.999980 \\
500 & 399.993601 & 400.000000 & 0.999984 \\
600  & 479.993601 & 480.000000 & 0.999987 \\
700  & 559.993601 & 560.000000 & 0.999989 \\
800  & 639.993601 & 640.000000 & 0.999990 \\
900  & 719.993601 & 720.000000 & 0.999991 \\
1000 & 799.993601 & 800.000000 & 0.999992 \\
\hline
\end{tabular}
\end{center}
\end{table}
Table~\ref{table:singlebiglam}, Table~\ref{table:single} and Table~\ref{table:singlesmalllam} show the cases where  %%@
the arrival rate at empty state is slightly greater than, equal to and slightly less than the service rate, %%@
respectively. In all cases, for sufficiently large value of the scaling factor $c$, the throughput of the server %%@
converges to the upper bound of the throughput. This result establishes our result that for a large system, the %%@
throughput of a server equals the minimum of arrival rate at empty state and service rate. Note that, when the system %%@
is not at empty state, the arrival rate is actually lower than the arrival rate of empty state. So, this result is %%@
counterintuitive and this insight  builds the foundation of our static request routing policy. Our routing policy %%@
chooses the surrogate server for a request by solving the optimization problem where the objective is to maximize the %%@
summation of the minimum of arrival rate at empty state and service rate for each of the servers of the CDN. As a %%@
result, our policy achieves the  upperbound of throughput for large system and so, it is asymptotically optimal for %%@
multiple server CDN.

\section{Conclusions}
\label{sec:conclusion}
We look at the big picture of content delivery network. Web based commercial CDNs compete with each other 
to grab the market share. The goal of the CEO is to maximize profit whereas  the 
design engineer looks at the system performance. We try to incorporate this two perspectives in one single framework %%@
and figure out
which design methodology maximizes monetary benefit. 
In our analysis of competition, we find that economy of scale effect is very significant here. So peering agreement %%@
among smaller CDNs may be a good 
idea to increase monetary benefits. Although we have analyzed the competition of CDNs using a model for two and three %%@
CDNs, same techniques can be used to analyze
competition among larger number of CDNs. However, with the increase of the number of CDNs the 
analysis becomes more and more tedious. Future works might include analyzing a more general model
having arbitrary number of competing CDNs.

 We also provide a static request routing policy which has been shown as asymptotically optimal. Although our model %%@
is developed in the context of a CDN, it is relevant to a 
 variety of contexts. For example, our approach can be used to determine the routing policy for 
 a website that maintains a number
 of mirrors in various places of the network and try to serve as many requests as 
 possible within a fixed user latency. 

%\section*{Acknowledgments}
\bibliographystyle{IEEE}
\bibliography{cdn}  
%\appendix
\end{document}